\documentclass[twocolumn,preprintnumbers,pra,amsmath,amssymb,superscriptaddress,showpacs]{revtex4}
\usepackage{epsfig}
\usepackage{graphicx}

\begin{document}

\title{Multipartite Nonlocality without Entanglement in Many Dimensions}

\author{J. Niset}
\affiliation{Quantum Information and Communication, Ecole
Polytechnique, CP 165, Universit\'e Libre de Bruxelles, 1050
Brussels, Belgium}

\author{N. J. Cerf}
\affiliation{Quantum Information and Communication, Ecole
Polytechnique, CP 165, Universit\'e Libre de Bruxelles, 1050
Brussels, Belgium}

\begin{abstract}
We present a generic method to construct a product basis exhibiting Nonlocality Without Entanglement with $n$ parties each holding a system of dimension at least $n-1$. This basis is generated via a quantum circuit made of control-Discrete Fourier Transform gates acting on the computational basis. The simplicity of our quantum circuit allows for an intuitive understanding of this new type of nonlocality. We also show how this circuit can be used to construct Unextendible Product Bases and their associated Bound Entangled States. To our knowledge, this is the first method which, given a general Hilbert space $\bigotimes_{i=1}^n {\cal H}_{d_i}$ with $d_i\le n-1$, makes it possible to construct (i) a basis exhibiting Nonlocality Without Entanglement, (ii) an Unextendible Product Basis, and (iii) a Bound Entangled state. 
\end{abstract}

\maketitle

\section{Introduction}
One of the most intriguing feature of quantum mechanics is entanglement.
As shown in the early 20th century by Einstein, Podolsky, and Rosen, 
quantum entanglement gives rise to nonlocality, namely the fact that
spatially separated systems may behave in a way that cannot be explained 
by any local theory \cite{EPR}. This effect, although it does not violate causality,
may nevertheless be verified experimentally, as originally discovered 
by Bell \cite{Bell}. More specifically, one can write
a Bell inequality that must be obeyed by any local realistic model but is
violated by quantum mechanics. 

Interestingly, there also exist other types of nonlocal behaviors, 
which go beyond entanglement. Inspired by an early work of Peres and Wootters
on a nonlocal effect manifested by correlated product states \cite{Peres}, 
Bennett {\it et al.} discovered a set of nine orthogonal product states 
in $3\otimes 3$ dimensions that cannot be perfectly distinguished 
if the two parties are restricted to Local Operations and Classical Communications (LOCC) \cite{Bennett}. This behavior was termed {\it Nonlocality without Entanglement} (NLWE) 
since we have a truly nonlocal behavior while
entanglement is used neither in the preparation of the states, nor
in the joint measurement that discriminates them perfectly. In contrast,
the model of Peres and Wootters concerned the joint measurement of a pair
of qubits prepared in a product (hence non-entangled) state via an observable admitting entangled eigenstates; they exhibited an example with correlated qubits where such a measurement gives more information than a separate 
LOCC-type measurement.

The NLWE behavior can be viewed as a new striking example of the nonequivalence between the concept of quantum entanglement and that of quantum nonlocality. It was know since Werner \cite{Werner} that entanglement does not necessarily imply nonlocality in the sense of producing data that are incompatible with local realism (entanglement $\nRightarrow$ nonlocality). This new type of nonlocality 
implies that the converse does not hold either (nonlocality $\nRightarrow$ entanglement). Note that nonlocality is not understood here as the incompatibility with local realism, but instead as the advantage of a joint measurement with respect to all LOCC strategies. More generally, NLWE thus raises the question: what kind of global operations can or cannot be performed using LOCC operations only? This question has attracted a lot of attention over the recent years as it underpins the use of entanglement as a resource in quantum information theory. 

Walgate {\it et al.} \cite{Hardy2} have shown that any two orthogonal quantum state, entangled or not, can be reliably distinguished using LOCC. Later, Walgate and Hardy \cite{Hardy} established the necessary and sufficient conditions for a general set of $2\times2$ quantum states to be locally distinguishable, and for a general set of $2\times n$ to be distinguished given an initial measurement of the qubit. These results reveal a fundamental {\em asymmetry} inherent to local distinguishability, which is conjectured to be at the origin of NLWE. For example, there exist sets of bipartite orthogonal product states that cannot be reliably distinguished locally if Bob is to go first and only one-way communication from him to Alice is allowed, but can be distinguished if Alice performs the first sequence of measurements and then shares her results with Bob. One such set is given by 
\begin{eqnarray} \label{set_g}
\nonumber \vert \Psi_1 \rangle &=& \vert 0 \rangle_a  \vert 0\rangle_b \\
\nonumber \vert \Psi_2 \rangle &=& \vert 0 \rangle_a \vert 1\rangle_b \\
\vert \Psi_3 \rangle &=& \vert 1 \rangle_a  \vert 0+1\rangle_b \\
\nonumber \vert \Psi_4 \rangle &=& \vert 1 \rangle_a  \vert 0-1\rangle_b 
\end{eqnarray}
where $\{\vert0\rangle, \vert1\rangle \}$ stands for the Computational Basis (CB) while $\{\vert0\pm1\rangle\}=\{\frac{1}{\sqrt{2}}(\vert0\rangle \pm \vert1\rangle)\}$ stands for the Dual Basis (DB). Groisman and Vaidman \cite{Vaidman} studied the {\it one-way indistinguishability} exhibited by this simple set, and used it to construct an alternative proof 
of the NLWE of the nine states found by Bennett {\it et al.} \cite{Bennett}.

In this paper, we further investigate this close connection between NLWE and one-way indistinguishability by introducing a 2-qubit quantum gate called the control-Hadamard (control-H).  This gate applies a Hadamard transform to one of the qubits conditioned on the other qubit being in the appropriate state of the CB, say $|1\rangle$. Not surprisingly,  this gate is closely connected to the set (\ref{set_g}) since it generates (\ref{set_g}) when applied on the four states of the CB. This control-Hadamard gate can be easily generalized to the $n$-partite case with $d$-dimensional systems: the Hadamard transform becomes a $d$-dimensional quantum Discrete Fourier Transform (DFT) applied to the $n$th system, while the control extends to the $n-1$ first parties. We will refer to this gate as a control-DFT$_d$ in the following. 

In Section~\ref{sect2}, we will show that this control-H gate provides us with a nice formalism to understand the mechanism of NLWE. Based on our observations of a known set exhibiting this phenomena, namely the so-called SHIFT ensemble introduced in \cite{Bennett}, we will derive 
in Section~\ref{sect3} a generic method to construct multipartite product bases exhibiting NLWE with systems of arbitrary dimension.  Finally, in Sections~\ref{sect4} and \ref{sect5}, we will exploit the fact that NLWE is connected to other peculiar quantum phenomena such as Unextendible Product Bases (UPB) \cite{UPB,UPB2} and Bound Entanglement (BE) \cite{horo1,horo2}, and will apply our method to generalize the construction of a large variety of UPBs and related BE states, a task that has proven to be extremely difficult in the past. \\

\section{Nonlocality Without Entanglement}
\label{sect2}

The SHIFT ensemble defined in \cite{Bennett} is the following set of 8 orthogonal three-qubit product states,
\begin{eqnarray} \label{SHIFT}
\nonumber \vert\Psi_1\rangle &=& \vert 0 \rangle_a\vert 0 \rangle_b\vert 0 \rangle_c \\
\nonumber\vert\Psi_2\rangle &=& \vert 0+1 \rangle_a\vert 0 \rangle_b\vert 1 \rangle_c \\
\nonumber\vert\Psi_3\rangle &=& \vert 0 \rangle_a\vert 1 \rangle_b\vert 0+1 \rangle_c \\
\nonumber\vert\Psi_4\rangle &=& \vert 0 \rangle_a\vert 1 \rangle_b\vert 0-1 \rangle_c \\
\nonumber\vert\Psi_5\rangle &=& \vert 1 \rangle_a\vert 0+1 \rangle_b\vert 0 \rangle_c \\
\nonumber\vert\Psi_6\rangle &=& \vert 0-1 \rangle_a\vert 0 \rangle_b\vert 1 \rangle_c \\
\nonumber\vert\Psi_7\rangle &=& \vert 1 \rangle_a\vert 0-1 \rangle_b\vert 0 \rangle_c \\
\vert\Psi_8\rangle &=& \vert 1 \rangle_a\vert 1 \rangle_b\vert 1 \rangle_c 
\end{eqnarray}
To understand why this set exhibits NLWE, consider the following game. An external party randomly chooses a number between 1 and 8 and accordingly prepares the corresponding quantum state $\vert\Psi_i\rangle$. He then sends the shares of this state to Alice, Bob, and Charles, who are located respectively at A, B, and C, and asks them to identify the state they have received, i.e., to perfectly determine the value of $i$. Note that Alice, Bob, and Charles know the precise form of the states of the set, but ignore which one has been prepared.  To distinguish between the possible states, the players may have recourse to LOCC only. This means that they are allowed to perform any sequence of local operations or measurements on their respective shares of the state and communicate their results to the other players through a classical channel, but cannot perform a joint measurement or communicate through a quantum channel. Indeed, if the players had access to joint operations, the problem would be trivial since the states are orthogonal, hence perfectly distinguishable. Surprisingly, it turns out that when the set of operations is restricted to LOCC, the players will never be able to perfectly distinguish between the 8 possible states \cite{Bennett}. Although the set is made of orthogonal product states, which can thus be prepared locally, it has the property of being {\it locally indistinguishable} (throughout this paper, we use indistinguishable to mean not perfectly distinguishable).\\

\begin{figure}[t] 
\includegraphics[width=0.7\linewidth]{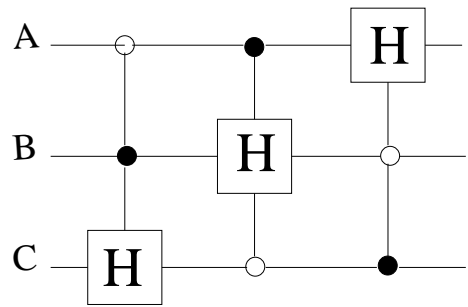}
\caption{Quantum circuit generating the SHIFT ensemble from the computational basis. The empty and filled circles correspond to a control condition of $\vert 0 \rangle$ and $\vert 1 \rangle$, respectively. For example, the first gate applies a Hadamard to Charles' qubit if Alice's qubit is $\vert 0 \rangle$ and Bob's qubit is $\vert 1 \rangle$.} \label{fig1}
\end{figure}

It is interesting to investigate how the joint measurement, which perfectly discriminates between the states of the set, can be 
implemented in terms of quantum logic gates. Formally, it is a projective
measurement based on the 8 projectors onto product states 
$\Pi_i=\vert \Psi_i \rangle \langle \Psi_i \vert$. 
Consider the 3-qubit unitary operation $U$, which transforms the 8 states
of the CB onto the states of the SHIFT ensemble, $\vert \Psi_i \rangle$. The knowledge of 
this joint unitary operation $U$ gives a simple strategy to perform the joint
measurement: it can be decomposed as the sequence of the joint unitary operation $U^\dagger$ followed by a local measurement by Alice, Bob, and Charles in the
computational basis. Interestingly, it turns out that the unitary $U$ can be implemented by a simple quantum circuit made of 3 identical control-Hadamard gates. As seen in Fig.\ref{fig1}, this gate is a tripartite gate which applies a Hadamard transform onto one of the qubits conditioned on the other two being in the appropriate product state $\vert 0 \rangle \vert 1 \rangle$. More precisely, in case the Hadamard acts on Charles' qubit, this control-H gate performs the operation 
\begin{displaymath}
\vert i\rangle_a\vert j \rangle_b \vert k \rangle_c\longrightarrow \left\{\begin{array}{ll}
\vert i\rangle_a\vert j \rangle_b H\vert k \rangle_c & \text{if} \ \ i=0 \wedge j=1 \\
\vert i\rangle_a\vert j \rangle_b \vert k \rangle_c & \text{otherwise}
             \end{array} \right.
\end{displaymath}
with $H|0\rangle=|0+1\rangle$ and $H|1\rangle=|0-1\rangle$.
Because of the cyclic control conditions, the 3 gates appearing in the circuit for $U$ are commuting; one can easily check that if the control conditions are satisfied for one of the gates, they cannot be satisfied for the two others. We will call this property \textit{exclusivity}. As an example, consider the first and second gates of the circuit. Expressing the Hadamard as $H=\exp(iG)$
with $G=(\pi/2)(\openone-H)$, these two gates can be written respectively as $\exp(iA)$ and $\exp(iB)$ with
\begin{eqnarray}
\nonumber  A &=& \frac{1}{4} \ (\openone+\sigma_z) \otimes (\openone-\sigma_z) \otimes G \\
\nonumber B &=& \frac{1}{4} \ (\openone-\sigma_z) \otimes G \otimes (\openone+\sigma_z)
\end{eqnarray}
We deduce from these expressions that $AB=BA=0$ since it contains
the product $(\openone+\sigma_z)(\openone-\sigma_z)=0$, which translates
the exclusivity property. Hence, $[A,B]=0$, and the Baker-Campbell-Hausdorff
formula gives
\begin{eqnarray*}
{\rm e}^{iA} {\rm e}^{iB} &=& {\rm e}^{i(A+B)} {\rm e}^{-[A,B]/2}\\
&=& {\rm e}^{i(A+B)} {\rm e}^{[A,B]/2}\\
&=& {\rm e}^{iB} {\rm e}^{iA} 
\end{eqnarray*}
which proves the commutation between 
the first and second gates of the circuit (the same reasoning holds for
any two gates). \\

\textit{Why does the circuit works?}
Consider first the ensemble constructed by applying only the first control-H (the one that acts on Charles' qubit) on the states of the CB. As shown in \cite{Hardy,Vaidman}, this ensemble is indistinguishable if Charles is forced to perform the first non-trivial step of the measurement strategy, or equivalently if he is restricted to one-way classical communication towards Alice and Bob. This is obvious as his share of the state could either be in the computational or in the dual basis, and any non-trivial measurement (one that will gain some information about one of these bases) will always irreversibly loose some information about the conjugate basis.  Of course, if Alice and Bob start while Charles is allowed to delay his measurement, then the set appears perfectly distinguishable. Alice and Bob should simply measure in the CB and then inform Charles about the basis he should use.  This is what Walgate and Hardy called the ``asymmetry of local distinguishability''. \\
\indent
Next, let us play the same game but using a quantum circuit made of the two first control-H gates (those acting on Charles' and Bob's qubits). 
The fact that the two gates are commuting (and exclusive) guarantees that no entanglement will be created when the sates of the CB are processed, i.e., the product states of the CB transform into another set of product states. This time, the ensemble appears indistinguishable to both Charles and Bob as their shares of the states are made of non-orthogonal, hence indistinguishable, states. This is obvious if we adopt a measurement point of view. The second gate tells us that Bob cannot start. But, since the gates commute, the second gate can be interchanged with the first, leading to the similar conclusion that Charles cannot start. Thus, in order to perfectly distinguish the states locally, neither Bob nor Charles may start. Again, if it is Alice who goes first, then the ensemble becomes locally distinguishable. She simply measures her share of the state in the CB: if she gets a $\vert 0 \rangle$ she knows that Bob's share should be measured in the CB, and the outcome of Bob's measurement determines which basis Charles should use. A $\vert 1 \rangle$ simply interchanges Charles' and Bob's roles. In short, this ensemble is locally indistinguishable if Charles or Bob are forced to start, but distinguishable if Alice goes first.\\

Now, consider the entire circuit of Fig.~\ref{fig1}, that is, the circuit made of the 3 control-H gates and the SHIFT ensemble that it generates. Following the two previous examples, we know that this ensemble is locally indistinguishable as each player sees an indistinguishable subset created by the Hadamard gate acting on his qubit (this gate can be placed last in the circuit). Consequently, in this case \textit{nobody} wants to start. Because in every LOCC strategy, \textit{someone} has to start, we get a simple picture of why the SHIFT ensemble exhibits NLWE.  This can be summarized as follows:
\begin{description}
\item 1. In every LOCC strategy, someone has to start
\item 2. The last gate implies that Alice cannot start
\item 3. The gates commute and can thus be interchanged
\item 4. By (2) and (3), nobody wants to start
\item 5. (1) and (4) are incompatible
\end{description}

\begin{figure}[t]
\includegraphics[width=0.9\linewidth]{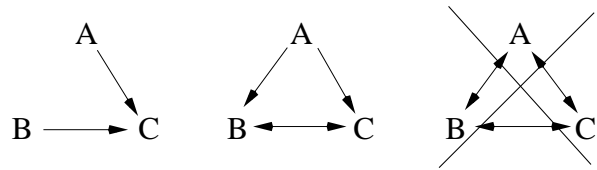}
\caption{In the first game (left), Charles needs information from both Alice and Bob to make the set distinguishable. In the second game (middle), Bob needs to receive information from Alice and transmit information to Charles, or the reverse, depending on Alice's measurement. In the third scenario (right), even if the players have access to all possible classical communication protocols, the set remains locally indistinguishable (NLWE).} \label{fig2}
\end{figure}

\noindent The three scenarios we have presented and the corresponding LOCC strategies can be nicely illustrated with the diagrams of Fig.~\ref{fig2}. The arrows represent the minimum amount of communication required to make the ensemble locally distinguishable. This intuitive picture will be translated into a rigorous proof in the next section.\\

\section{n-partite Nonlocality Without Entanglement in arbitrary dimension} 
\label{sect3}

With the intuition gained from our circuit, we see that what is really at issue in NLWE is not the kind of LOCC protocols employed by the parties, nor the content of their communication, but rather the asymmetry of local distinguishability encapsulated in the states themselves. Indeed, it is because each player does not know which of the two conjugated basis he should use to measure his share of the state (the set is indistinguishable from his point of view) and because all players are in this same situation (the gates commute) that the ensemble is locally indistinguishable as a whole. All we have used so far is thus the possibility, for each player, to have a state belonging to two conjugate bases. We may therefore extend our construction to systems of arbitrary dimension instead of qubits.
Given the central role played by the control-H gate, which creates this local indistinguishability, we may replace it with a $d$-dimensional quantum Discrete Fourier Transform (DFT$_d$). This yields a simple strategy to construct an ensemble made of orthogonal tripartite product states of arbitrary dimensions that exhibits NLWE. Note that by changing the control conditions of the gates while maintaining their exclusivity, we can define a whole family of NLWE ensembles with equivalent properties. Furthermore, there is no need to restrict the circuit to a tripartite scenario. Knowing that the key ingredient is a sequence of control-DFT gates that are exclusive (hence commuting), we can further generalize the method and increase the number of parties. The quantum circuit will now be made of $n$ control-DFT gates, one acting on each player, and we require these gates to be exclusive to make sure that the resulting ensemble is made of product states. This imposes some constraint on the dimensions of the players' shares. A simple way to satisfy this exclusivity condition is to require that each player has a Hilbert space large enough to accommodate $n-1$ gates with a different control state. We can therefore state the following sufficient condition
\begin{equation} \label{dn}
d_j \geq n-1 
\end{equation}
where $d_j$ is the dimension of ${\cal H}_j$, the Hilbert space of player $j$, and $n$ is the total number of players. We thus have established a generic method to construct some $n$-partite ensemble of product states exhibiting NLWE using systems of arbitrary dimension: \\

\textit{Statement: If we have $n\geq3$ parties working in respective Hilbert spaces ${\cal H}_j$ of dimension $d_j\geq n-1$, a quantum circuit can be defined, based on $n$ control-DFT gates, which generates a set $\{\vert \Psi_i \rangle\}$ made of $\prod_j d_j$ orthogonal product states that form a basis of $\bigotimes_j {\cal H}_j$ and exhibit Nonlocality Without Entanglement.}\\

\begin{figure}[t]
\includegraphics[width=0.8\linewidth]{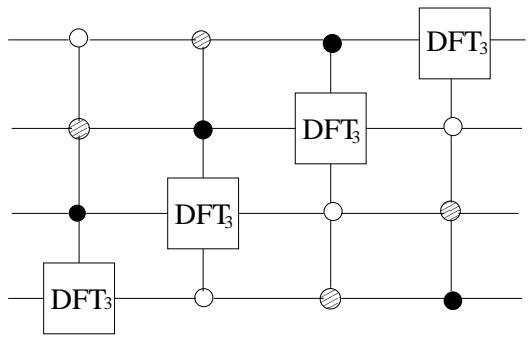}
\caption{Nonlocality without entanglement in  dimension $3\otimes3\otimes3\otimes3$. The first gate applies a Discrete Fourier Transform in dimension $3$ to Damian's qutrit if Alice's state is a $\vert 0\rangle$, Bob's state a $\vert 1\rangle$ and Charles' state a $\vert 2\rangle$.} \label{fig3}
\end{figure}

\textit{Proof:}
Let us denote the CB of player $j$ by $\{ |0\rangle,\cdots |d_j -1\rangle\}$. We consider the ensemble generated from the CB of all players by applying the unitary 
\begin{equation}\label{SHIFTd}
U=\prod_{j=1}^n {\rm e}^{i A_j}
\end{equation}
with $A_1=\bigotimes_{j=1}^{n-1} |j-1\rangle\langle j-1| \otimes G$ and $A_{2,3,\cdots n}$ are cyclic permutations of $A_1$. In the simplest case, each party has a dimension $d_j=n-1$, saturating relation (\ref{dn}), but this is not necessary for the proof to hold. As an example, we show in Fig.~\ref{fig3} the circuit generating this simplest ensemble exhibiting NLWE for a quadripartite scenario in which all parties hold a qutrit, i.e., in a Hilbert space of total dimension $3\otimes3\otimes3\otimes3$. 

Let us proof now that the ensemble generated by (\ref{SHIFTd}) exhibits NLWE. Suppose Alice has a share of dimension $d$ and performs the first step of the measurement procedure. We will show that, under the simple constraint of not allowing her operation to lead to a situation in which it has become impossible in principle to perfectly distinguish between the initial states $\vert \Psi_i\rangle$, then she cannot gain any information. First, let us rewrite the states of the ensemble as $\vert \Psi_i \rangle = \vert \phi_i\rangle_A\otimes\vert\varphi_i\rangle_B$ where $\vert \phi_i \rangle$ is Alice's share and $\vert\varphi_i\rangle_B$ is the state held by all the other players. We describe Alice's measurement in two stages: first, her share of the state and the measuring device evolve unitarily under the action of some unitary operator $U_A$; second, some outcome $k$ is read out. The unitary evolution of Alice's share and the measuring device can be described by:
\begin{equation} \label{eqd1}
U_A: \vert\phi_i\rangle_A\vert A \rangle \longrightarrow \sum_k \alpha_{ik} \vert \omega_{ik} \rangle
\end{equation}
where $\vert A \rangle$ is the initial state of the measuring device, and $\vert \omega_{ik} \rangle$ is the joint state of Alice's share and the measuring device corresponding to a particular outcome k. Without restriction, we can choose $\alpha_{ik}$ to be real and non-negative. Importantly, the states
$\vert \omega_{i,k} \rangle$ with different $k$ must be orthogonal as they correspond to different outcomes of the macroscopic measuring device. 
Note that for each $k$, only $2d$ couples $\{\alpha_{ik}, \vert \omega_{ik}\rangle\}$ are necessary to completely describe the unitary evolution $U_A$ since Alice only sees $2d$ distinct states $\vert\phi_i\rangle_A$ ($d$ states of the CB and $d$ states of the DB). We therefore redefine the unitary 
evolution $U_A$ via the $2d$ relations
\begin{eqnarray}
\nonumber\vert m\rangle_A\otimes\vert A \rangle &\overset{U_A}{\longrightarrow}& \sum_k \alpha_{m,k}\vert \omega_{m,k}\rangle \hspace{17mm} \\
\Vert m\rangle_A\otimes\vert A \rangle &\overset{U_A}{\longrightarrow}& \sum_k \alpha'_{m,k}\vert \omega'_{m,k}\rangle \hspace{10mm} 
\end{eqnarray}
with $k$ labeling the different outcomes, $m=0,\ldots,d-1$, and $\Vert m \rangle$ denoting the states of the DB.

Next, we must impose some constraints on Alice's possible operations. On the one hand, she has to distinguish between the $d$ states that are in the DB at her side since these states appear identical to the other parties. On the other hand, there are also states that are in the CB but are only distinguishable by Alice as they appear nonorthogonal to the other parties [e.g., the states $|\Psi_3\rangle$ and $|\Psi_8\rangle$ of the ensemble (\ref{SHIFT}), or the states $\vert 0\rangle_A \vert 2\rangle_{B} \vert 2\rangle_{C} \vert 2\rangle_{D}$ and $\vert 1\rangle_A \vert 2\rangle_{B} \vert 2\rangle_{C} \vert 2\rangle_{D}$ in the example of Fig.~\ref{fig3}]. For those states, which she sees as orthogonal, she must maintain perfect distinguishability whatever action she performs and for every value of the outcome $k$. In other words, her measurement must either distinguish these states outright or leave them orthogonal, i.e., we must impose for every $k$ and for all $m\neq n = 0,1, ... ,d-1$
\begin{eqnarray} \label{orthd}
\nonumber \alpha_{m,k}\alpha_{n,k}\langle \omega_{m,k}\vert \omega_{n,k} \rangle &=& 0 \hspace{10mm}  \\
\alpha'_{m,k}\alpha'_{n,k}\langle \omega'_{m,k}\vert \omega'_{n,k} \rangle &=& 0 \hspace{10mm} 
\end{eqnarray}
In addition to these $d(d-1)$ constraints, we also need to consider the relations between the possible initial states at Alice's site. More precisely, we know that the CB and DB are related by the quantum Fourier Transform :
\begin{eqnarray} \label{reld1}
\nonumber \Vert n \rangle  &=& \frac{1}{\sqrt{d}} \sum_{l=0}^{d-1} \exp(i\frac{2\pi}{d}nl)\ \vert l \rangle \\
\vert m \rangle  &=&  \frac{1}{\sqrt{d}} \sum_{l=0}^{d-1} \exp(-i\frac{2\pi}{d}ml)\ \Vert l \rangle 
\end{eqnarray}
The unitary evolution $U_A$ must conserve these relations, and, since $\vert \omega_{m,k} \rangle$ with different $k$ are orthogonal, we can write $2d$ relations of the form 
\begin{eqnarray} \label{rr}
\nonumber \alpha'_{lk}\vert \omega'_{lk} \rangle &=& \frac{1}{\sqrt{d}} \sum_{j=0}^{d-1} \exp(i\frac{2\pi}{d}jl)\ \alpha_{jk}\vert \omega_{jk}\rangle 
 \\
\alpha_{lk}\vert \omega_{lk} \rangle &=& \frac{1}{\sqrt{d}} \sum_{j=0}^{d-1} \exp(-i\frac{2\pi}{d}jl)\ \alpha'_{jk}\vert \omega'_{jk}\rangle \end{eqnarray}
for each outcome $k$, with $l=0,\ldots,d-1$. Consider the last $d$ relations of (\ref{rr}). If we take the scalar product of two of them and reorganize appropriately the different terms of the sums, we obtain
\begin{eqnarray} \label{equd2}
\nonumber \alpha_{lk}\alpha_{l'k}\langle \omega_{lk}\vert \omega_{l'k} \rangle=\frac{1}{d}\Big( \sum_{j=0}^{d-1} \exp\big(-i\frac{2\pi}{d}j(l-l')\big) \alpha_{jk}^{'2} \\
+ 2 \sum_{j=0}^{d-1} \sum_{j'>j} \cos\big(\frac{2\pi}{d}(lj-l'j')\big)\alpha'_{jk}\alpha'_{j'k}\langle \omega'_{j'k}\vert \omega'_{jk} \rangle \Big) 
\end{eqnarray}
If we now choose $l=l'$, it follows
\begin{eqnarray} 
\alpha_{lk}^2 &=&\frac{1}{d}\Big( \sum_{j=0}^{d-1}\alpha_{jk}^{'2}\\ 
\nonumber&+& 2 \sum_{j=0}^{d-1} \sum_{j'>j} \cos\big(\frac{2\pi}{d}l(j-j')\big)\alpha'_{jk}\alpha'_{j'k}\langle \omega'_{j'k}\vert \omega'_{jk} \rangle \Big) 
\end{eqnarray}
By the second condition of (\ref{orthd}), all the terms of the second sum of the right-hand side are trivially equal to zero. Since the remaining term does not depend on the value of $l$, we conclude that, for each value of the outcome $k$, all the $\alpha_{lk}$'s are equal. Similarly,
from the other $d$ conjugated relations of (\ref{rr}), we conclude that for all $k$ the $\alpha'_{lk}$'s must be equal and equal to the $\alpha_{lk}$'s. It follows that if $k$ is a possible outcome for one particular initial state $\vert \psi_i\rangle$ , then $k$ is a possible outcome for all the initial states. In addition, for all such outcomes $k$ the distinguishability condition (\ref{orthd}) becomes the true orthogonality condition 
\begin{eqnarray} 
 \nonumber \langle \omega_{mk}\vert \omega_{nk} \rangle &=& 0 \hspace{10mm} \forall m\neq n \in\{ 0,1,\cdots d-1 \}\\
\langle \omega'_{mk}\vert \omega'_{nk} \rangle &=& 0 \hspace{10mm} \forall m\neq n \in \{0,1,\cdots d-1\} 
\end{eqnarray}
To summarize, after a measurement procedure which produced the outcome $k$, the set of possible initial states of Alice's share together with the measuring device have evolved into a set of states which is isomorphic to the initial set, i.e., no information can be gained and communicated to the other players from the value of $k$. Thus, in order to perfectly distinguish between the states, Alice cannot start. In view of the commutation of the gates, similar arguments can be used to show that the other players face the same dilemma of either gaining some useful information at the cost of irreversibly loosing perfect distinguishability, or not gaining any information at all. This completes the proof. $\square$ \\

\begin{figure}[t]
\includegraphics[width=0.8\linewidth]{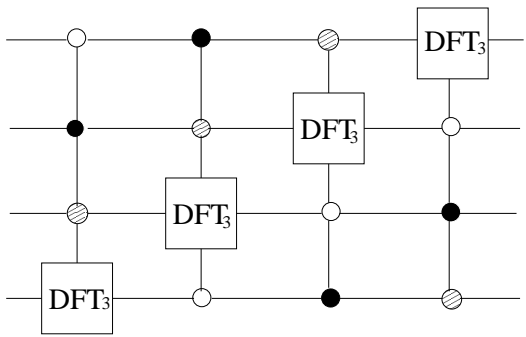}
\caption{By concatenating this circuit to the one of Fig.~\ref{fig3}, one gets another ensemble exhibiting NLWE in dimension $3\otimes3\otimes3\otimes3$. The first gate applies a Discrete Fourier Transform in dimension $3$ to Damian's qutrit if Alice's state is a $\vert 0\rangle$, Bob's state a $\vert 2\rangle$ and Charles' state a $\vert 1\rangle$.} \label{fig4}
\end{figure}

To conclude this section, it is worth noticing that the sets we have constructed so far are not the most general ones. Indeed, for $d>2$ the size of the local Hilbert space allows for the introduction of more gates without loosing the exclusivity (or commutation) condition. For instance, a gate triggered by $\vert0\rangle_A \vert2\rangle_B \vert1\rangle_C$ and acting on Damian's qutrit can be added to the circuit of Fig.~\ref{fig3} while conserving the commutation condition. The set constructed with the circuit of Fig.~\ref{fig3} supplemented with this gate and cyclic permutations (as shown in Fig.~\ref{fig4}) exhibits NLWE while it is qualitatively distinct from the set of Fig.~\ref{fig3}. In particular, it has more shares in the dual bases, so we conjecture that it should be ``more nonlocal'' in this respect. Nevertheless, for the rest of this paper we will only consider sets constructed with the minimum number of gates $n$. As a last comment, let us note that although our method \textit{as it is} fails to create bipartite NLWE, the $9$ bipartite qutrit states (``domino'' states) of \cite{Bennett} do fit into our picture of commuting control gates. We can associate to this set a quantum circuit made of 4 control-gates, two for each player, where the exclusivity of the gates requires the Fourier Transform to act on 2-dimensional subspaces of ${\cal H}_A$ and ${\cal H}_B$.  \\

\section{N-Partite Unextendible Product Bases}
\label{sect4}

\textit{Definition: Consider a n-partite quantum system belonging to
${\cal H}=\otimes_{i=1}^n {\cal H}_i$, with the local Hilbert spaces of respective dimensions $d_i$. An Unextendible Product Basis is an incomplete orthogonal Product Basis whose complementary subspace contains no product state.}\\

Unextendible Product Bases (UPB) are known to be closely connected to NLWE. More precisely, Bennett et al. \cite{UPB} showed that the members of a UPB cannot be perfectly distinguished if one is restricted to LOCC, i.e. they exhibit NLWE (see also \cite{derinaldis}). A necessary and sufficient condition for extendibility of a product basis is also known, and has been used to construct a UPB from a complete product basis exhibiting NLWE. This method was used, for example, to construct the following UPB 
\begin{eqnarray} \label{SHIFTUPB}
\nonumber\vert\Psi_4\rangle &=& \vert 0 \rangle_a\vert 1 \rangle_b\vert 0-1 \rangle_c \\
\nonumber\vert\Psi_6\rangle &=& \vert 0-1 \rangle_a\vert 0 \rangle_b\vert 1 \rangle_c \\
\nonumber\vert\Psi_7\rangle &=& \vert 1 \rangle_a\vert 0-1 \rangle_b\vert 0 \rangle_c \\
\vert\Psi_\mathrm{st}\rangle &=& \vert 0+1 \rangle_a\vert 0+1 \rangle_b\vert 0+1 \rangle_c 
\end{eqnarray}
from the SHIFT ensemble \cite{UPB,UPB2}. The extra state $|\Psi_{st}\rangle$ here stands for the ``stopper'' state, as explained later on.
Since we have found in the previous Section a systematic way of creating product bases exhibiting NLWE, we can further exploit these ideas and give a systematic protocol to construct UPBs based on quantum circuits such as those of Fig.~\ref{fig1} and \ref{fig3}. Although the protocol we present here most probably does not account for the construction of all possible UPBs, it nevertheless can be used to construct a family of UPBs in a large variety of scenarios. \\

We will illustrate the construction with a simple example. Consider the quadripartite circuit of Fig.~\ref{fig3} and suppose Alice, Bob, Charles, and Damian hold systems  of dimension $d_a$, $d_b$, $d_c$, and $d_d$, respectively. We have shown that this circuit generates an ensemble $\{ \vert \Psi_i \rangle\}$ made of $d_a d_b d_c d_d$ orthogonal product states exhibiting NLWE. We construct an UPB out of $\{\vert \Psi_i \rangle \}$
by extracting from it the $d_a-1$ states in which Alice's share is any state
of the DB except the last one, the $d_b-1$ states in which Bob's share is any state of the DB except the last one, the $d_c-1$ states in which Charles' share is any state of the DB except the last one, and the $d_d-1$ states in which Damian's share is any state of the DB except the last one.
We complete these $\sum_{i=1}^4(d_i-1)$ states by adding a proper \textit{stopper} state, so as to force the unextendibility of the set. 
Note that the number of states $m$ in a UPB is known to verify 
the condition \cite{UPB}
\begin{equation}
m\geq\sum_{i=1}^n(d_i-1)+1
\end{equation}
so that the above construction yields a minimal UPB.
 
More specifically, the UPB consists of the states
\begin{eqnarray}\label{UPB}
\nonumber \vert\Psi_{1}^0\rangle&=&\vert 0 \rangle_a \vert 1 \rangle_b \vert 2 \rangle_c  F\vert 0 \rangle_d\\
\nonumber &\vdots& \\
\nonumber \vert\Psi_1^{d_d-2}\rangle&=&\vert 0 \rangle_a \vert 1 \rangle_b \vert 2 \rangle_c  F\vert d_d-2 \rangle_d\\
\nonumber \\
\nonumber \vert\Psi_2^{0}\rangle&=&\vert 1 \rangle_a \vert 2 \rangle_b F\vert 0 \rangle_c\vert 0 \rangle_d\\
\nonumber &\vdots& \\
\nonumber\vert \Psi_2^{d_c-2}\rangle&=&\vert 1 \rangle_a \vert 2 \rangle_b F\vert d_c-2 \rangle_c \vert 0 \rangle_d\\
\nonumber\\
\nonumber\vert \Psi_3^{0\rangle}&=&\vert 2 \rangle_a  F\vert 0 \rangle_b \vert 0 \rangle_c  \vert 1 \rangle_d\\
\nonumber &\vdots& \\
 \nonumber \vert\Psi_3^{d_b-2}\rangle&=&\vert 2 \rangle_a  F\vert d_b-2 \rangle_b \vert 0 \rangle_c \vert 1 \rangle_d\\
\nonumber\\
 \vert\Psi_4^{0}\rangle&=&F\vert 0 \rangle_a \vert 0 \rangle_b \vert 1 \rangle_c  \vert 2 \rangle_d\\
\nonumber &\vdots& \\
\nonumber \vert\Psi_4^{d_a-2}\rangle&=&F\vert d_a-2 \rangle_a \vert 0 \rangle_b \vert 1 \rangle_c \vert 2 \rangle_d\\
\nonumber\\
\nonumber\vert\Psi_\mathrm{st}\rangle&=& F\vert d_a-1 \rangle_a  F\vert d_b-1 \rangle_b  F\vert d_c-1 \rangle  F\vert d_d-1 \rangle
\end{eqnarray}
where $F\vert i \rangle$ is the Discrete Fourier Transform of $\vert i \rangle$. To understand why this set is unextendible, suppose we want to add a new product state that is orthogonal to it. Clearly, because of the $d_a-1$ states that are in the DB for Alice together with the stopper sate, we cannot find a state orthogonal to Alice's share (this subset of $d_a$ states span her entire subspace). So, we should try to look for a product state that is orthogonal to this set within Bob's, Charles' or Damian's subspaces. But the same argument as for Alice's subspace can be applied as well. Hence no product state can be found that is orthogonal to all the states, and the set is unextendible. 

\section{N-partite Bound Entangled States}
\label{sect5}

In addition to being locally indistinguishable, UPBs are also connected to another important quantum property, namely Bound Entanglement. A Bound Entangled (BE) state is an entangled mixed state from which no pure entanglement can be distilled \cite{horo1,horo2}. The role of bound entanglement in nature, and its yet-to-find possible use for quantum information processing has attracted a lot of attention. Although the construction of bound entangled states has proven to be a difficult task, it has been realized that the state corresponding to the uniform mixture on the subspace orthogonal to a UPB $\{\vert \tilde{\psi}_i \rangle , i=1,\cdots m\}$, namely
\begin{equation} \label{be}
\hat{\rho}=\frac{1}{D-m}\big( 1-\sum_{i=1}^m \vert \tilde{\psi}_i \rangle\langle \tilde{\psi}_i \vert \big)
\end{equation}
is a bound entangled state \cite{UPB}, where $D$ is the total dimension of
the Hilbert space. This is one of the only known generic method to construct bound entangled states. The quantum circuit we have introduced provides us with a simple strategy to construct a large number of UPBs, hence it also provides us with a simple method to construct a large number of BE states.
Consider again our quadripartite UPB, as defined in Eq.~(\ref{UPB}). By definition the space complementary to the UPB contains no product states, hence $\hat{\rho}$ is entangled. We can use the partial transposition to show that every partitioning of the parties is PPT: indeed, the identity is invariant under partial transposition and the product states $\vert \tilde{\psi}_i \rangle$ of the UPB are mapped onto other product states. Thus, no entanglement can be distilled across any bipartite cut. In addition, note that if some pure global entanglement could nevertheless be distilled, it could be used to create entanglement across a bipartite cut; hence no entanglement at all can be distilled and the state is indeed bound entangled. \\

In the special case where the exclusivity condition (\ref{dn}) is saturated, that is, when we have $n=d+1$ parties holding each a system of dimension $d$, the BE state produced by this method has another interesting property, namely that the entanglement across any $d\otimes d^{d}$ cut is zero (this is stronger than the nondistillability of bipartite entanglement across any $d\otimes d^{d}$ cut). This result was already derived for the special case of the SHIFT ensemble \cite{UPB}, i.e., for $d=2$, and we will show here how to use a similar argument to prove that it is the case for any value of $d$. \\

To show that the entanglement across the cut A/BC...E is zero, we explicitly separate Alice from the other parties and rewrite the $d^2$ states of the UPB as
\begin{eqnarray}
\nonumber \vert\Psi_{1,i}\rangle &=&\vert 0\rangle \vert a_i \rangle,  \hspace{10.1mm} \vert a_i\rangle=\vert 1,2,\ldots,d-2,F(i) \rangle  \\
\nonumber \vert\Psi_{2,i}\rangle &=&\vert 1\rangle \vert b_i \rangle,  \hspace{10.75mm}  \vert b_i\rangle=\vert 2,\ldots,d-2,F(i),0 \rangle \\
\nonumber &&\hspace{5mm}\vdots \\
\nonumber \vert\Psi_{d,i}\rangle &=&\vert d-1\rangle \vert e_i \rangle,  \hspace{5mm} \vert e_i\rangle=\vert F(i),0,\ldots,d-3 \rangle \\
\nonumber \vert\Psi_{d+1,i}\rangle &=&\vert F(i)\rangle \vert f\rangle, \hspace{7mm} \vert f\rangle\ =\vert 0,\ldots,d-1 \rangle \\
\nonumber\vert \Psi_{st}\rangle &=&\vert F(d-1)\rangle \vert g \rangle,  \hspace{0.5mm} \vert g\rangle\ =\vert F(d-1),\ldots,F(d-1) \rangle 
\end{eqnarray}
where $i=0,\ldots,d-2$, and $F(i)$ means $F\vert i \rangle$. 
Next note that $\vert f \rangle$ and $\vert g \rangle$ span a Hilbert space $S=$span($f$,$g$) of dimension $2$, and that all the states in this space are orthogonal to $\{\vert a_i\rangle,\vert b_i\rangle,\ldots,\vert e_i\rangle\}$. We thus define the Hilbert space $S'=\mathcal{H}(d^{d})/S$ of dimension $d^d-2$ such that (i) all the states in this space are, by construction, orthogonal to $\vert f \rangle$ and $\vert g \rangle$; and (ii) all the states $\{\vert a_i\rangle,\vert b_i\rangle,\ldots,\vert e_i\rangle\}$ are in $S'$. We can therefore find an ensemble of $d^d-d-1$ orthogonal vectors $\vert a_k^\perp \rangle$ such that every $\vert a_k^\perp\rangle$ belongs to $S'$ and is orthogonal to all the $\vert a_i\rangle$, i.e. $\{\vert a_i\rangle,\vert a_k^\perp\rangle \}$ is an orthogonal basis of $S'$. We repeat that procedure for the $\{\vert b_i\rangle\}$ and define $d^d-d-1$ vectors $\vert b_k^\perp \rangle$ in $S'$, until we have defined the last $d^d-d-1$ vectors $\vert e_k^\perp\rangle$ associated to the states $\{\vert e_i\rangle\}$.  In addition, we can also define $\vert f^\bot \rangle$ and $\vert g^\bot \rangle$ in $S$, orthogonal to $\vert f \rangle$ and $\vert g \rangle$ respectively. We can now use all these new vectors to complete the original UPB and make it a full $d^{d+1}$-dimensional product basis between A and BC...E. This is done by adding the $d(d^d-d-1)+(d-1)+1=d^{d+1}-d^2$ new states $ \{\vert 0\rangle \vert a_k^\perp \rangle, \vert 1\rangle \vert b_k^\perp \rangle,\ldots,\vert d-1\rangle \vert e_k^\perp \rangle,\vert F(i)\rangle \vert f^\perp\rangle,\vert F(d-1)\rangle \vert g^\perp \rangle\}$. This shows that with respect to the cut A and BC...E, the set is completable by product states and the mixed state $\hat{\rho}$ is therefore not entangled. Because the state is symmetric, this argument also applies to the other $d\otimes d^{d}$ splits which prove that our generic bound entangled state contains no entanglement across any such cuts.

\section{conclusions}

We have developed a systematic approach to Nonlocality Without Entanglement based on observations of the SHIFT ensemble introduced by Bennett {\it et al.} \cite{Bennett}. Our approach, centered on commuting control-Hadamard gates, connects NLWE to a simple quantum circuit consisting of such gates. This method gives an intuitive understanding of NLWE, and allows for a generalization both in the number of parties and in the dimension of their spaces. We therefore derive the first method to construct $n$-partite $d$-dimensional Product Bases exhibiting NLWE. 
For example, in Fig.~\ref{fig3}, we display the quantum circuit generating the set exhibiting NLWE with 4 qutrits; to our knowledge, it is the first example of NLWE in $3\otimes3\otimes3\otimes3$. 

The ingredients of our method are very general, and incorporate also the construction of the other known examples of NLWE introduced in \cite{Bennett}. In contrary to the original work of Ref. \cite{Bennett}, we did not attempt to place a bound on the mutual information attainable through LOCC, but rather adopted the strategy of \cite{Vaidman} and restricted ourselves to proving that such a nontrivial bound exists. However, a strategy to calculate this bound is needed as it would allow us to compare the degree of nonlocality exhibited by the different ensembles we can construct with this method, so it is an interesting challenge for future work. Because NLWE is connected to other quantum peculiarities such as Unextendible Product Bases and Bound Entanglement, our method can be adapted to construct $n$-partite high-dimensional UPBs and their associated BE states, a task that has proven to be difficult in the past. 


\section{acknowledgments}
We acknowledge financial support from the Communaut\'e Fran\c caise
de Belgique under grant ARC 00/05-251, from the IUAP programme of the 
Belgian government under grant V-18, and from the EU under projects 
COVAQIAL and SECOQC. J.N. also acknowledges support from the Belgian
FRIA foundation.

\end{document}